\documentclass[12pt,a4j]{article}
\usepackage[dvips]{graphicx}
\usepackage{enumerate}
\usepackage{amsmath}

\begin{document}\title{Dispersionless Limits of  Integrable Generalized Heisenberg Ferromagnet Equations}
\author{Zhaidary Myrzakulova\footnote{Email: zhrmyrzakulova@gmail.com}, \, Gulgassyl Nugmanova\footnote{Email: nugmanovagn@gmail.com}, \, Kuralay Yesmakhanova\footnote{Email: krmyesmakhanova@gmail.com}, \\ and  Ratbay Myrzakulov\footnote{Email: rmyrzakulov@gmail.com}\\
Eurasian International Center for Theoretical Physics and \\  Department of General \& Theoretical Physics, \\ Eurasian National University,
Astana, 010008, Kazakhstan
}
\date{}
\maketitle
\begin{abstract}
This paper is a continuation of our previous work in which we studied a dispersionless limits of some integrable spin systems. Now, we shall present dispersionless limits of some integrable generalized Heisenberg ferromagnet equations
\end{abstract}

\section{Introduction} 
One of classical equations integrable through inverse scattering transform is the famous Heisenberg ferromagnet equation (HFE) \cite{l77}-\cite{Takhtajan}
\begin{eqnarray}
{\bf A}_{t}={\bf A}\wedge {\bf A}_{xx}, \label{1} 
\end{eqnarray}
where 
\begin{eqnarray}
{\bf A}=(A_{1}, A_{2}, A_{3}), \quad {\bf A}^{2}=1. 
\end{eqnarray}
The Lax representation (LR) of the HFE reads as
\begin{eqnarray}
\Psi_{x}&=&U_{1}\Psi, \\
\Psi_{t}&=&V_{1}\Psi,
\end{eqnarray}
where
\begin{eqnarray}
U_{1}&=&-i\lambda A, \\
V_{2}&=&-2i\lambda^{2}A+0.5i\lambda [A,A_{x}].
\end{eqnarray}
Here
\begin{eqnarray}
A= \left(\begin{array}{cc}
A_{3} & A^{-} \\ 
 A^{+} & -A_{3}\end{array}\right), \quad A^{\pm}=A_{1}\pm iA_{2}.
\end{eqnarray}
After discovery the integrability of the HFE were constructed several class integrable and nonintegrable generalized  HFE in 1+1 and 2+1 dimensions (see e.g. \cite{Gerdjikov1}-\cite{R4} and references therein). Integrable dispersionless equations play important role in modern physics and mathematics. In this context,    dispersionless limits of some integrable spin systems   were found \cite{z1}-\cite{z2}. In the present paper, we intend to consider  dispersionless limits of some integrable generalized Heisenberg ferromagnet equations, e.g.,  the Gerdjikov-Mikhailov-Valchev equation (GMVE). 

The paper is organized as follows. Next section introduces the so-called M-? equation and its Lax representation (LR).  In section 3, we present the dispersionless limit of this equation. The dispersionless limit of the GMVE is presented in section 4. Section 5 contains some further discussion and remarks.  

\section{M-CI equation}

Consider the following M-CI equation (which is a spin system with the self-consistent potential) 
\begin{eqnarray}
{\bf A}_{t}+{\bf N}\wedge {\bf A}_{xx}+J(2u{\bf A}_{x}+u_{x}{\bf A})+
(B_{2}{\bf A})_{x}&=&0,  \\
u_{t}+2uu_{x}+(B_{2}u)_{x}-(\ln {A_{3}})_{xxx}-\{[(\ln{A_{3}})_{x}]^{2}\}_{x}&=& 0, 
\end{eqnarray}
where   ${\bf A}=(A_{1}, A_{2}, A_{3})$ is a unit spin vector that is ${\bf A}^{2}=\delta(A_{1}^{2}+A_{2}^{2})+A_{3}^{2}=1$  ($\delta^{2}=1$), $u=\chi_{x}$ is a real function (scalar potential), ${\bf N}=(0,0,-1)$, $J=diag(0,0,1)$, $A^{\pm}=A_{1}\pm iA_{2}$   and 
\begin{equation}
 B_{2}=\delta(A_{1x}A_{2}-A_{1}A_{2x})-A_{3}^{2}\chi_{x}=0.5i\delta(A^{+}_{x}A^{-}-A^{+}A^{-}_{x})-A_{3}^{2}u.
\end{equation}
In components, the M-CI equation takes the form
\begin{eqnarray}
A_{1t}+A_{2xx}+(B_{2}A_{1})_{x}  &=&  0, \\
A_{2t}-A_{1xx}+(B_{2}A_{2})_{x}  &=&  0, \\
A_{3t}+2A_{3x}\chi_{x}+A_{3}\chi_{xx}+(B_{2}A_{3})_{x}  &=&  0, \\
\chi_{t}-(\ln {A_{3}})_{xx}-[(\ln{A_{3}})_{x}]^{2}+\chi_{x}^{2}+B_{2}\chi_{x}&=& 0, 
\end{eqnarray}
or
\begin{eqnarray}
A_{1t}+A_{2xx}+(B_{2}A_{1})_{x}  &=&  0, \\
A_{2t}-A_{1xx}+(B_{2}A_{2})_{x}  &=&  0, \\
A_{3t}+2A_{3x}u+A_{3}u_{x}+(B_{2}A_{3})_{x}  &=&  0, \\
u_{t}-(\ln {A_{3}})_{xxx}-\{[(\ln{A_{3}})_{x}]^{2}\}_{x}+2uu_{x}+(B_{2}u)_{x}&=& 0. 
\end{eqnarray}
We can write the M-CI  equation also in the following form
\begin{eqnarray}
iA^{+}_{t}+A_{xx}^{+}+i(B_{2}A^{+})_{x}  &=&  0, \\
A_{3t}+2A_{3x}\chi_{x}+A_{3}\chi_{xx}+(B_{2}A_{3})_{x}  &=&  0, \\
\chi_{t}-(\ln {A_{3}})_{xx}-\{[(\ln{A_{3}})_{x}]^{2}\}+\chi_{x}^{2}+B_{2}\chi_{x}&=& 0, 
\end{eqnarray}
or
\begin{eqnarray}
iA^{+}_{t}+A_{xx}^{+}+i(B_{2}A^{+})_{x}  &=&  0, \\
A_{3t}+2A_{3x}u+A_{3}u_{x}+(B_{2}A_{3})_{x}  &=&  0, \\
u_{t}-(\ln {A_{3}})_{xxx}-\{[(\ln{A_{3}})_{x}]^{2}\}_{x}+2uu_{x}+(B_{2}u)_{x}&=& 0, 
\end{eqnarray}
where $A^{\pm}=A_{1}\pm iA_{2}$ and 
\begin{equation}
  B_{2}=0.5i\delta(A^{+}_{x}A^{-}-A^{+}A^{-}_{x})-A_{3}^{2}u=0.5i\delta(A^{+}_{x}A^{-}-A^{+}A^{-}_{x})-A_{3}^{2}\chi_{x}.
\end{equation}

The M-CI equation is integrable. It LR has the form
\begin{eqnarray}
\Psi_{x}&=&U_{1}\Psi, \\
\Psi_{t}&=&V_{1}\Psi,
\end{eqnarray}
where
\begin{eqnarray}
U_{1}&=&-i\lambda K, \\
V_{2}&=&i\lambda^{2}K_{2}+i\lambda K_{1}.
\end{eqnarray}
Here
\begin{eqnarray}
K &=& \left(\begin{array}{ccc}
0 & A^{+} & A_{3}e^{iu}\\ 
\delta A^{-} & 0 & 0 \\
A_{3}e^{-iu} & 0 & 0\end{array}\right), \quad K_2  = \left(\begin{array}{ccc}
- 1/3 & 0 & 0 \\ 0 & 2/3 - \delta|A^{+}|^2 & -\delta A^{-}A_{3}e^{iu} \\
0 & -  A^{+}A_{3}e^{-iu} & 2/3 -u^{2}
\end{array}\right),\\
K_1 & = & \left(\begin{array}{ccc}
0 & a & b\\ 
\delta a^* & 0 & 0 \\
b^* & 0 & 0\end{array}\right),\quad
a = - i A^{+}_{x} - iAA^{+},\quad 
b = - i (A_{3x}+iu_{x}A_{3}+AA_{3})e^{iu},   \end{eqnarray}
where
\begin{eqnarray}
A=\delta A^{+}A^{-}_{x} + A_{3}(A_{3x}-iu_{x}A_{3}).
\end{eqnarray}
We note that to get this LR of the M-CI equation, we have used the corresponding LR of the GMVE \cite{Gerdjikov1}.  

\section{M-CII equation}

Now we want to derive the dispersionless limit of the M-CI  equation, using,  for example, the equations (19)-(21). Consider the transformation
\begin{equation}
 A^{+}=\sqrt{w}e^{\frac{i}{\epsilon}S},
\end{equation}
where 
\begin{equation}
 A_{3}=\sqrt{1-\delta w}.
\end{equation}
Substituting the expression (33) into (19)-(21), in the dispersionless limit, we  get the following M-CII  equation
\begin{eqnarray}
w_{t}+2(S_{x}w)_{x}-[(A^{2}_{3}\chi_{x}+\delta w S_{x})\sqrt{w}]_{x} &=&  0, \\
S_{t}+S^{2}_{x}-S_{x}[A^{2}_{3}\chi_{x}+\delta w S_{x}]_{x}  &=& 0, \\
\chi_{t}+\chi^{2}_{x}-\chi_{x}[A_{3}^{2}\chi_{x}+\delta wS_{x}]&=&0
\end{eqnarray}
or
\begin{eqnarray}
w_{t}+2(S_{x}w)_{x}-[(A^{2}_{3}u+\delta w S_{x})\sqrt{w}]_{x} &=&  0, \\
S_{t}+S^{2}_{x}-S_{x}[A^{2}_{3}u+\delta w S_{x}]_{x}  &=& 0, \\
u_{t}+2uu_{x}-[u(A_{3}^{2}u+\delta wS_{x})]_{x}&=&0.
\end{eqnarray}
We can write this M-CII equation also in the following form
\begin{eqnarray}
w_{t}+2(vw)_{x}-[(A^{2}_{3}u+\delta w v)\sqrt{w}]_{x} &=&  0, \\
v_{t}+2vv_{x}-[v(A^{2}_{3}u+\delta w v)]_{x}  &=& 0, \\
u_{t}+2uu_{x}-[u(A_{3}^{2}u+\delta wv)]_{x}&=&0.
\end{eqnarray}
These three forms of the M-CII  equation are the equivalent forms of the dispersionless limit of the M-CI equation. Note that in the above equations,  $A_{3}$ is given by the expression (34).

\section{Dispersionless limit of GMVE}
Consider the GMVE  \cite{Gerdjikov1}
\begin{eqnarray}
iq_{1t}+q_{1xx}+(Aq_{1})_{x}&=&0  \\
iq_{2t}+q_{2xx}+(Aq_{2})_{x}&=&0, 
\end{eqnarray}
where
\begin{equation}
\delta|q_{1}|^{2}+|q_{2}|^{2}=1
\end{equation}
and
\begin{eqnarray}
A=\epsilon{q_{1}}\bar{q}_{1x}+q_{2}\bar{q}_{2x}.
\end{eqnarray}
Let us now introduce the Madelung tranformations
\begin{eqnarray}
q_{1}&=&\sqrt{u_{1}}e^{\frac{i}{\epsilon}S_{1}},\\
q_{2}&=&\sqrt{u_{2}}e^{\frac{i}{\epsilon}S_{2}}.
\end{eqnarray}
Then from (46) we have 
\begin{equation}
\delta u{_{1}}+u_{2}=1,
\end{equation}
\begin{equation}
u_{2}=1-\delta u{_{1}}.
\end{equation}
Substituting the expressions (48)-(49) into the GMVE (44)-(45), in the dispersionless limit, we obtain the following M-CIII equation
\begin{eqnarray}
u_{1t}+2[S_{1x}u_{1}]_{x}-2\sqrt{u_{1}}([2]\sqrt{u_{1}})_{x}&=&0,\\
u_{2t}+2[S_{2x}u_{2}]_{x}-2\sqrt{u_{2}}([2]\sqrt{u_{2}})_{x}&=&0,\\
S_{1t}+{S^{2}_{1x}}-[2]{S_{1x}}&=&0,\\
S_{2t}+{S^{2}_{2x}}-[2]{S_{2x}}&=&0,
\end{eqnarray}
or
\begin{eqnarray}
u_{1t}+2(u_{1}v_{1})_{x}-2\sqrt{u_{1}}([2]\sqrt{u_{1}})_{x}&=&0,\\
u_{2t}+2(u_{2}v_{2})_{x}-2\sqrt{u_{2}}([2]\sqrt{u_{2}})_{x}&=&0,\\
v_{1t}+({v^{2}_1}-[2]v_1)_{x}&=&0,\\
v_{2t}+({v^{2}_2}-[2]v_2)_{x}&=&0,
\end{eqnarray}
where $v=S_{x}$ and 
\begin{eqnarray}
\delta\frac{u_{1x}}{2}+\frac{u_{2x}}{2}=[1]&=&0,\\
SS_{1x}u_{1}+S_{2x}u_{2}&=&[2].
\end{eqnarray}
Thus the M-CIII equation is the dispersionless limit of the GMVE. The LR of the M-CIII equation  reads as 
\begin{eqnarray}
p-\lambda\left[ \frac{\delta u_{1}}{p-v_{1}}+\frac{u_{2}}{p-v_{2}}\right]&=&0\\
p_{t}+\lambda\left\{\frac{\delta}{p-v_{1}}\left[v_{1}-\delta u_{1}v_{1}-u_{2}v_{2}\right]+\frac{u_{2}}{p-v_{2}}\left[v_{2}-\delta u_{1}v_{1}-u_{2}v_{2}\right]\right\}_x&=&0
\end{eqnarray}
or
\begin{eqnarray}
p-\lambda\left[ \frac{\delta u_{1}}{p-v_{1}}+\frac{u_{2}}{p-v_{2}}\right]&=&0, \\ 
p_{t}+\lambda\left[\frac{\delta u_{1}\left(v_{1}-B\right)}{p-v_{1}}+\frac{u_{2}\left(v_{2}-B\right)}{p-v_{2}}\right]_x&=&0, 
\end{eqnarray}
where
\begin{eqnarray}
B=\delta u_{1}v_{1}+u_{2}v_{2}.
\end{eqnarray}

\section{Conclusion}

In this paper,  we have considered some generalized Heisenberg ferromagnet equations which are integrable by IST method. The dispersionless limits of these equations were found. Also we have shown that the M-CI  equation is equivalent to the GMVE.

\section{Acknowledgements}
This work was supported in part by the Ministry of Edication  and Science of Kazakhstan under
grant 0118RK00935  as well as by grant  0118RK00693.

\end{document}